\documentstyle[titlepage,preprint,aps,floats,epsfig]{revtex}
\newcommand{\beq}{\begin{equation}}
\newcommand{\eeq}{\end{equation}}
\newcommand{\eq}[1]{eq.(\ref{#1})}
\begin{document}
\draft
\preprint{}
\tighten
\title {Three-Loop Radiative-Recoil
Corrections to Hyperfine Splitting Generated by One-Loop
Fermion Factors}
\medskip
\author {Michael I. Eides
\thanks{E-mail address:  eides@pa.uky.edu, eides@thd.pnpi.spb.ru}}
\address{Department of Physics and Astronomy,
University of Kentucky,
Lexington, KY 40506, USA\\
and
Petersburg Nuclear Physics Institute,
Gatchina, St.Petersburg 188300, Russia}
\author{Howard Grotch\thanks{E-mail address: hgrotch@uky.edu}}
\address{Department of Physics and Astronomy, University of Kentucky,
Lexington, KY 40506, USA}
\author{Valery A. Shelyuto \thanks{E-mail address:
shelyuto@vniim.ru}}
\address{D. I.  Mendeleev Institute of Metrology,
St.Petersburg 198005, Russia}

\maketitle

\begin{abstract}
We consider three-loop radiative-recoil corrections to
hyperfine splitting in muonium generated by diagrams with  one-loop
radiative photon insertions both in the electron and muon lines. An
analytic result for these nonlogarithmic corrections of order
$\alpha(Z^2\alpha)(Z\alpha)(m/M)\widetilde E_F$ is obtained. This
result constitutes a next step in the implementation of the program of
reduction of the theoretical uncertainty of hyperfine splitting below
10 Hz.
\end{abstract}


\section{Introduction}

Three-loop radiative-recoil corrections to hyperfine splitting in
muonium are enhanced by the large logarithm cubed of the electron-muon
mass ratio \cite{es0} (see, also review \cite{egs01r}). The leading
logarithm cubed contribution is generated by the graphs with insertions
of the electron one-loop polarization operators in the two-photon
exchange graphs. It may be obtained almost without any calculations by
substituting the effective charge $\alpha(M)$ in the leading recoil
correction of order $(Z\alpha)(m/M)\widetilde E_F$, and expanding the
resulting expression in a power series in $\alpha$. Calculation of the
logarithm squared term of order $\alpha^2(Z\alpha)(m/M)\widetilde E_F$
is more challenging \cite{eks89}. Different graphs generate logarithm
squared terms, and all such contributions were obtained a long time ago
\cite{es0,eks89,kes90}. The sum of the logarithm cubed and  logarithm
squared terms is given by the expression \footnote{We define the Fermi
energy  as

\beq      \label{baremuonfermi}
\widetilde{E}_{F}=\frac{16}{3}Z^4\alpha^2
\frac{m}{M} \left(\frac{m_r}{m}\right)^{3}ch\:R_{\infty},
\eeq

\noindent
where $m$ and $M$ are the electron and muon masses, $m_r$ is the
reduced mass of the electron-muon system, $\alpha$ is the fine structure
constant, $c$ is the velocity of light, $h$ is the Planck constant,
$R_{\infty}$ is the Rydberg constant, and $Z$ is the nucleus charge in
terms of the electron charge ($Z=1$ for muonium).  The Fermi energy
$\widetilde{E}_{F}$ does not include the muon anomalous magnetic moment
$a_\mu$ which does not factorize in the case of recoil corrections, and
should be considered on the same grounds as other corrections to
hyperfine splitting.}

\begin{equation}
\Delta
E=\left(-\frac{4}{3}\ln^3\frac{M}{m}+\frac{4}{3}\ln^2\frac{M}{m}\right)
\frac{\alpha^2(Z\alpha)}{\pi^3}\frac{m}{M} {\widetilde E}_F.
\end{equation}

Due to recent experimental and theoretical progress, single-logarithmic
and nonlogarithmic contributions of orders
$\alpha^2(Z\alpha)(m/M)\widetilde E_F$ and
$\alpha(Z^2\alpha)(Z\alpha)(m/M)\widetilde E_F$ to hyperfine splitting
in muonium are now also phenomenologically relevant. Numerous sets of
gauge invariant diagrams generate single-logarithmic and nonlogarithmic
contributions.

\begin{figure}[ht]
\centerline{\epsfig{file=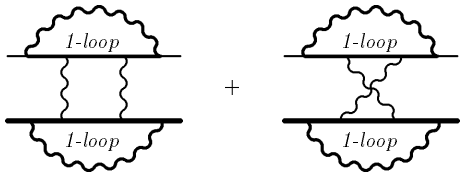,height=2.5cm}}
\vspace{0.5cm}
\caption{Diagrams with two fermion factors}
\label{radrecdiag}
\end{figure}

\begin{figure}[ht]
\centerline{\epsfig{file=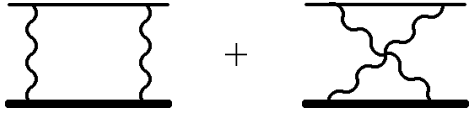,height=1.5cm}}
\vspace{0.5cm}
\caption{Diagrams with two-photon exchanges}
\label{twophothfsfig}
\end{figure}

Below we consider three-loop radiative-recoil corrections to hyperfine
splitting in muonium generated by the diagrams in Fig.\
\ref{radrecdiag}. These diagrams are obtained from the skeleton
diagrams in  Fig.\ \ref{twophothfsfig} by making all one-loop
radiative photon insertions both in the electron and muon lines. The
two-loop radiative-recoil corrections  generated by all one-loop
radiative photon insertions only in the electron or only in the muon
line were calculated earlier (see, e. g., review in \cite{egs01r}).
The leading recoil correction of order $Z\alpha(m/M)\widetilde E_F$ is
generated by the skeleton diagrams with two exchanged photons in Fig.\
\ref{twophothfsfig}. The characteristic loop momenta in the skeleton
diagrams are larger than the electron mass, and therefore the leading
recoil correction to hyperfine splitting may be calculated in the
scattering approximation, ignoring the wave function momenta of order
$mZ\alpha$ (see, e. g., \cite{egs01r}). It was obtained a long time ago
\cite{arn,fm,newsalp}

\beq              \label{arnnewcsalp}
\Delta E_{rec}=-\frac{3mM}{M^2-m^2}\frac{Z\alpha}{\pi}\ln\frac{M}{m}
~{\widetilde E}_F.
\eeq

\noindent
Note that this correction is proportional to the logarithm of the
electron-muon mass ratio, and quite remarkably it turns out that the
logarithmic contribution is a complete result, there is no
nonlogarithmic contribution of this order.

The radiative insertions  can  only increase the characteristic
integration momenta in the diagrams  in Fig.\ \ref{twophothfsfig} and
hence the scattering approximation remains valid for calculation of
two- and three-loop radiative-recoil corrections. The two-loop
radiative-recoil corrections of order $\alpha(Z\alpha)(m/M)\widetilde
E_F$ generated by the one-loop radiative photon insertions in the
electron line are logarithmic in the electron-muon mass ratio. Since
the leading recoil correction of order $Z\alpha(m/M)\widetilde E_F$ is
linear in the logarithm of the large mass ratio, one could expect that
the correction of order $\alpha(Z\alpha)(m/M)\widetilde E_F$ is
proportional to the logarithm squared. This does not happen and the
logarithm squared contributions cancel as was first discovered in
\cite{cp} by direct calculation. The simplest way to understand this
cancellation is to recall that in the Landau gauge radiative insertions
in the electron line are nonlogarithmic \cite{blp}, and hence, being
gauge invariant, the sum of these insertions is nonlogarithmic in any
gauge.  The single-logarithmic term of order
$\alpha(Z\alpha)(m/M)\widetilde E_F$ was obtained in \cite{ty}, and the
nonlogarithmic terms were calculated numerically in \cite{sty} and
analytically in \cite{eks5}

\beq         \label{electronaslog}
\Delta
E=\biggl[\frac{15}{4}\ln\frac{M}{m}
+6\zeta(3)+3\pi^2\ln2+\frac{\pi^2}{2}+\frac{17}{8}
\biggr]\frac{\alpha(Z\alpha)}{\pi^2}\frac{m}{M}~{\widetilde E}_F.
\eeq

\noindent
One more feature of the calculations in \cite{sty,eks5} deserves to be
mentioned. One-loop radiative insertions in the electron line include
the terms connected with the one-loop anomalous magnetic moment.  These
terms have different low energy behavior in comparison with all other
terms in the dressed electron line and could in principle compilcate
calculation of the radiative-recoil corrections. However, as was
discovered in \cite{sty,eks5} the terms connected with the one-loop
anomalous magnetic moment do not give any contribution at all to the
radiative-recoil corrections of order $\alpha(Z\alpha)(m/M)\widetilde
E_F$. Finally, let us mention that numerically the nonlogarithmic part
of the correction of order $\alpha(Z\alpha)(m/M)\widetilde E_F$ is
rather large, of order $\pi^2$, which is  just what one should expect
for the constants accompanying the large logarithm.

The two-loop radiative-recoil corrections of order
$(Z^2\alpha)(Z\alpha)(m/M)\widetilde E_F$ are generated by all one-loop
radiative photon insertions only in the muon line in the diagrams in
Fig.\ \ref{twophothfsfig}, and were obtained in \cite{sty,eks88}

\beq
\Delta
E=\biggl[\frac{9}{2}\zeta(3)-3\pi^2\ln2+\frac{39}{8}\biggr]
\frac{(Z^2\alpha)(Z\alpha)}{\pi^2}\frac{m}{M}~{\widetilde
E}_F.
\eeq

\noindent
Two features of these corrections deserve to be mentioned. First,
radiative insertions in the muon line do not generate logarithmic terms
at all, as can be understood with the help of the generalized low
energy theorem \cite{ty,eksann2}.  Second, just as in the case of the
insertions in the electron line the terms connected with the muon
anomalous magnetic moment do not give any contribution to the
radiative-recoil corrections of order
$(Z^2\alpha)(Z\alpha)(m/M)\widetilde E_F$.

In this work we analytically calculate three-loop radiative-recoil
corrections to hyperfine splitting in muonium generated by the diagrams
in Fig.\ \ref{radrecdiag} with all one-loop radiative photon insertions
both in the electron and muon lines. We show that these
corrections are nonlogarithmic and unlike the case of the
radiative-recoil corrections of orders $\alpha(Z\alpha)(m/M)\widetilde E_F$
and $(Z^2\alpha)(Z\alpha)(m/M)\widetilde E_F$ the one-loop anomalous
magnetic moments of both particles give nonvanishing contributions to
the correction under investigation.

\section{Gauge Invariant Representation for Radiative
Corrections}\label{logcanc}

Let us consider the general structure of the radiative-recoil
corrections in Fig.\ \ref{radrecdiag}. To this end it is convenient to
introduce the one-loop fermion factor $L_{\mu \nu}(k)$ as a sum of
the diagrams in Fig.\ \ref{fermionfactor}.  In terms of the electron
and muon factors the radiative-recoil contribution to hyperfine
splitting generated by the ladder and crossed-ladder diagrams in Fig.\
\ref{radrecdiag} has the form

\begin{equation} \label{elfactmuonfact}
\Delta E =-\frac{3}{8}\frac{(Z\alpha)mM}{\pi}~\widetilde E_F
\int \frac{d^4 k}{i \pi^2(k^2+i0)^2}\biggl[ L_{\mu\nu}^{(e)}(k) +
L_{\nu\mu}^{(e)}(-k) \biggr]  L_{\mu\nu}^{(\mu)}(-k).
\eeq

\noindent
The sum of the electron factors
$C^{(e)}_{\mu\nu}(k)\equiv L^{(e)}_{\mu\nu}(k)+L^{(e)}_{\nu\mu}(-k)$
which enters \eq{elfactmuonfact} for the radiative corrections is
just the gauge invariant Compton scattering amplitude for a virtual
photon, and satisfies the identity
$k^\mu C^{(e)}_{\mu\nu}(k)=0$. The electron
Compton amplitude is invariant under the substitution $k\to -k$ and
$\mu\to \nu$, and hence, we can substitute the muon Compton
amplitude instead of the muon factor in the integral in
\eq{elfactmuonfact} $L_{\mu\nu}^{(\mu)}(-k)\to
[L^{(\mu)}_{\mu\nu}(-k)+L^{(\mu)}_{\nu\mu}(k)]/2\equiv
C^{(\mu)}_{\mu\nu}(-k)/2$ obtaining a more symmetric expression for the
energy shift

\begin{equation} \label{elfactmuonfactsym}
\Delta E =-\frac{3}{16}\frac{(Z\alpha)mM}{\pi}~\widetilde E_F
\int \frac{d^4 k}{i \pi^2k^4}C^{(e)}_{\mu\nu}(k)C^{(\mu)}_{\mu\nu}(-k).
\eeq

\begin{figure}[ht]
\centerline{\epsfig{file=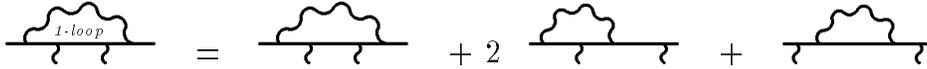,height=1cm}}
\vspace{0.5cm}
\caption{Fermion factor}
\label{fermionfactor}
\end{figure}

To simplify further calculations we represent the electron and muon
Compton amplitudes in \eq{elfactmuonfactsym} as sums of two gauge
invariant terms (we write the formula only for the electron, and the
respective expression for the muon is obtained by the substitution
$m\to M$, $\alpha\to Z^2\alpha$)

\beq  \label{sumrepr}
C^{(e)}_{\mu\nu}(k)=
C^{(e,a)}_{\mu\nu}(k) + C^{(e,b)}_{\mu\nu}(k),
\eeq

\noindent
where

\beq   \label{elfactamm}
C^{(e,a)}_{\mu \nu}(k)=\frac{\alpha}{2\pi}
\frac{1}{2m}\Biggl[\frac{\sigma_{\mu\rho}k^{\rho}({\hat p} - {\hat k}
+ m)\gamma_{\nu} ~+~ \gamma_{\mu} ({\hat p} - {\hat k} + m)
\sigma_{\nu\rho}(-k^{\rho})}{k^2 - 2mk_0}
\eeq
\[
+\frac{\sigma_{\nu\rho}(-k^{\rho})({\hat p} + {\hat k} + m)\gamma_{\mu}
~+~ \gamma_{\nu} ({\hat p} + {\hat k} + m) \sigma_{\mu\rho}k^{\rho}}
{k^2 + 2mk_0}\Biggr]
\]

\noindent
corresponds to the anomalous magnetic moment,
and  $C^{(e,b)}_{\mu \nu}(k)$ includes all other terms.

It is easy to check directly that  $C^{(e,a)}_{\mu\nu}(k)$, and hence,
$C^{(e,b)}_{\mu \nu}(k)$ are gauge invariant. The
breakdown in \eq{sumrepr} is helpful because $C^{(e,a)}_{\mu
\nu}(k)$, and $C^{(e,b)}_{\mu\nu}(k)$ have different behavior at
small photon momenta $k$. As we will see below this different low
energy behavior determines the structure of integrals for the
contributions to hyperfine splitting.

We can further simplify the amplitude $C^{(e,a)}_{\mu \nu}(k)$,
preserving only the terms which contribute to hyperfine splitting.
The simplified expression (still satisfying the Ward identity
$k^\mu C^{(e,a)}_{\mu\nu}(k)=0$) has the form

\beq \label{elfactammval}
C^{(a)}_{\mu\nu}(k)=\frac{\alpha}{\pi}\frac{k^2 \,\gamma_{\mu}
{\hat k} \gamma_{\nu}
~-~ k^2 \, k_0 \gamma_{\mu} \gamma_{\nu}
~+~ k_0 \,\bigl(k_{\mu} {\hat k} \gamma_{\nu}
+ \gamma_{\mu} {\hat k}k_{\nu}\bigl)}{k^4 - 4m^2k_0^2}.
\eeq

In terms of the representation in \eq{sumrepr} the contribution to
hyperfine splitting in \eq{elfactmuonfact} can be written as a sum of
three gauge invariant terms

\begin{equation} \label{generalcontr}
\Delta E =-\frac{3}{16}\frac{(Z\alpha)mM}{\pi}~\widetilde E_F
\int \frac{d^4 k}{i \pi^2k^4}
\Bigl[C^{(e,a)}_{\mu \nu}(k)C^{(\mu,a)}_{\mu
\nu}(-k)
+C^{(e,b)}_{\mu \nu}(k)C^{(\mu,a)}_{\mu
\nu}(-k)
\eeq
\[
+C^{(e)}_{\mu \nu}(k)C^{(\mu,b)}_{\mu \nu}(-k)\Bigr]
\equiv \Delta E^{I}+  \Delta E^{II} + \Delta E^{III}.
\]

\noindent
It is important to note that we know in advance that there is no
logarithm of the mass ratio in the sum of all contributions in
\eq{generalcontr}. Such a logarithm can only arise from the
integration region $m<k<M$, where the electron factor is in the
asymptotic regime. The asymptotic expression for the electron factor
was calculated e.g., in \cite{eks89}, and contains only the skeleton
spinor structure $\gamma_\mu\hat k\gamma_\nu$. On the other hand, all
terms in the muon factor except the term with the muon anomalous
magnetic moment are, in this integration region, additionally
suppressed by an extra factor $k^2/M^2$ in comparison with the
logarithmic skeleton integral, and thus cannot produce a
logarithmic contribution. As to the term with the muon anomalous
magnetic moment, its contribution to the recoil integral vanishes
identically due to its spinor structure, see, e. g., \cite{eksann2}.

\section{Two Anomalous Magnetic Moments}

Let us start our calculation with the term $\Delta E^{I}$ connected
with the product of two one-loop anomalous magnetic moments.
Projecting the spinor structures of the fermion
factors, written in the form of \eq{elfactamm} or \eq{elfactammval}, on
hyperfine splitting we obtain

\beq
\Delta E^{I}=\frac{\alpha(Z^2\alpha)(Z\alpha)mM}{\pi^3}
~\widetilde E_F \int \frac{d^4 k}{i
\pi^2k^4}\frac{k^2(k^4+4k^2k_0^2+k_0^4)}{(k^4-4k_0^2m^2)
(k^4-4k_0^2M^2)},
\eeq

\noindent
or after the Wick rotation and transition to the spherical
coordinates $k_0=k\cos\theta$, $|{\bf k}|=k\sin\theta$

\beq
\Delta E^{I}=
\frac{2\alpha(Z^2\alpha)(Z\alpha)mM}{\pi^4}
~\widetilde E_F \int_0^\infty dk^2\int_0^\pi d\theta\sin^2\theta
\frac{(1+4\cos^2\theta+\cos^4\theta)}{(k^2+4m^2
\cos^2\theta)
(k^2+4M^2\cos^2\theta)}.
\eeq

\noindent
Calculating the angular integral we discover that the remaining
momentum integral diverges like $dk^2/k^3$. This divergence indicates
the existence of the nonrecoil correction of order $\alpha(Z^2\alpha)$,
which is of lower order in $Z\alpha$. It is connected with the
one-photon exchange, and is well known. We subtract this power
divergence and, after the subtraction, obtain a convergent integral,
which can be easily calculated

\beq  \label{new1}
\Delta E^{I}
=\frac{9}{8}\frac{\alpha(Z^2\alpha)(Z\alpha)}{\pi^3}
\frac{mM}{M^2-m^2}\ln\frac{M}{m}~\widetilde E_F .
\eeq

\section{Subtracted Electron Factor and the Muon
Anomalous Magnetic Moment}

The second contribution in \eq{generalcontr} arises from the product
of the subtracted electron factor and the muon anomalous magnetic
moment, and we write it in the form

\beq \label{eiiiint}
\Delta E^{II}=-\frac{3}{8}\frac{(Z\alpha)mM}{\pi}~\widetilde E_F
\int \frac{d^4 k}{i \pi^2k^4}L^{(e,b)}_{\mu \nu}(k)C^{(\mu,a)}_{\mu
\nu}(-k).
\eeq

\noindent
The muon Compton amplitude is gauge invariant and satisfies
the Ward identity $k^\mu C^{(\mu,a)}_{\mu\nu}(k)=0$. Therefore,
we can omit all terms in the subtracted electron factor which
are proportional to $k_\mu$. This means that we can use the expression
for the subtracted electron factor from \cite{beks,egs98}, where all
terms proportional to $k_\mu$ are thrown away. We represent this
electron factor as a sum of seven terms
$L^{(e,b)}_{\mu\nu}(k)=\sum_1^7L^{(i)}_{\mu\nu}(k)$, which are

\begin{equation}             \label{fermionfact}
L^{(1)}_{\mu \nu}(k)+L^{(2)}_{\mu \nu}(k) = \frac{\alpha}{4\pi}
\langle \gamma_{\mu} \hat {k} \gamma_{\nu}\rangle _{(e)}
\int_0^1 dx  \int_0^x
\frac{dy}{(-k^2 + 2mbk_0 + m^2a^2)^3}
\left(c_{1} m^2 {\bf k}^2+c_{2}  k^4  \right),
\end{equation}
\[
L^{(3)}_{\mu \nu}(k)+L^{(4)}_{\mu \nu}(k) = \frac{\alpha}{4\pi}
\langle \gamma_{\mu} \hat {k} \gamma_{\nu}\rangle _{(e)}
\int_0^1 dx  \int_0^x
\frac{dy}{(-k^2 + 2mbk_0 + m^2a^2)^2}
\left(c_{3} k^2+2c_{4}m k_0 \right),
\]
\[
L^{(5)}_{\mu \nu}(k)+L^{(6)}_{\mu \nu}(k) = \frac{\alpha}{4\pi}
\langle \gamma_{\mu} \gamma_{\nu}\rangle _{(e)}
\int_0^1 dx  \int_0^x
\frac{dy}{(-k^2 + 2mbk_0 + m^2a^2)^2}
\left(c_{5} m k^2+2c_{6}  k^2  k_0 \right),
\]
\[
L^{(7)}_{\mu \nu}(k) = \frac{\alpha}{4\pi}
\langle \gamma_{\mu} \gamma_{\nu}\rangle _{(e)}
\int_0^1 dx  \int_0^x
\frac{dy}{-k^2 + 2mbk_0 + m^2a^2}
\left(c_{7} \frac{k^2}{m}\right),
\]

\noindent
where $k^2=k_0^2-{\bf k}^2$. Each term in the electron factor
corresponds to the respective coefficient function $c_i$ , and explicit
expressions for these coefficient functions are collected in Table
\ref{table1}.  We preserve in the electron factor only the spinor
structures $\langle \gamma_{\mu} \hat {k} \gamma_{\nu}\rangle _{(e)}$
and $\langle \gamma_{\mu} \gamma_{\nu}\rangle _{(e)}$ relevant for
hyperfine splitting, and the projection on hyperfine splitting is
understood. Auxiliary functions of the  Feynman parameters $a(x,y)$ and
$b(x,y)$ are defined by the relationships

\begin{equation}
a^2 = \frac{x^2}{y(1 - y)}, \qquad b= \frac{1 - x}{1 - y}.
\end{equation}

\noindent
The explicit expression for the muon factor can be obtained from the
expression for the electron factor by the substitutions $m\to M$ and
$\alpha\to Z^2\alpha$.

\begin{table}
\caption{Coefficients in the Fermion Factor}
\begin{tabular}{ll}
\\
$\mbox{c}_{1}$& $\frac{16}{y(1-y)^3}
\Big[~(1-x)(x-3y)~-~2y\ln{x}~\Big]~$
\\ \\ \tableline
\\
$\mbox{c}_{2}$      &   $\frac{4}{y(1-y)^3}
\Big[~-(1-x)(x - y - 2y^2 /x)~+~2(x - 4y + 4y^2 /x)  \ln{x}~\Big]~$
\\ \\ \tableline
\\
$\mbox{c}_{3}$     &   $\frac{1}{y(1-y)^2}
\Big[~1 - 6x - 2x^2 -(y/x) (26 - 6y/x - 37x - 2x^2
+ 12xy + 16 \ln{x})~\Big]~$
\\ \\ \tableline
\\
$\mbox{c}_{4}$   & $\frac{1}{y(1-y)^2}
\Big(~2x - 4x^2 - 5y + 7xy~ \Big)~$
\\ \\ \tableline
\\
$\mbox{c}_{5}$   & $\frac{1}{y(1-y)^2}
\Big(~ 6x - 3x^2 - 8y + 2xy ~\Big)~$
\\ \\ \tableline
\\
$\mbox{c}_{6}$   & $-  \frac{(1-x)^2(x - y)}{x^2(1-y)^2}~$
\\ \\ \tableline
\\
$\mbox{c}_{7}$  & $2  \frac{1-x}{x}~$
\\ \\
\end{tabular}
\label{table1}
\end{table}

Taking projection on the hyperfine splitting and contracting the
Lorentz indices, we obtain the integral for the  contribution to the
hyperfine splitting as

\begin{equation}      \label{electrcontr}
\Delta E^{II} = \alpha (Z\alpha)(Z^2\alpha) ~\widetilde E_F~  \frac{1}{8
\pi^3 \mu}  \int_0^1 {dx} \int_0^x {dy} \int \frac{d^4 k}{i \pi^2}
\frac{1}{(k^2 + i0)^2(k^4 - \mu^{-2}k^2_0)}
\end{equation}
\[
\Bigl\{ (6k^2_0 - 2{\bf k}^2)
\Bigl[\frac{c_{1}  {\bf k}^2 + c_{2}  (k^2)^2}
{(-k^2 + 2bk_0 + a^2)^3}
+\frac{ c_{3}  k^2 + c_{4}  2k_0}
{(-k^2 + 2bk_0 + a^2)^2} \Bigr]
\]
\[
- (6+2\frac{{\bf k}^2}{k^2})k_0  \Bigl[
\frac{c_{5}  k^2 + c_{6}  k^2  2k_0}
{(-k^2 + 2bk_0 + a^2)^2}
+ \frac{c_{7}  k^2}
{-k^2 + 2bk_0 + a^2} \Bigr] \Bigr\},
\]

\noindent
where $\mu=m/(2M)$ and we rescaled the integration momentum, so that
now it is measured in units of the electron mass.

The analytic calculation of the integrals in \eq{electrcontr} is one of
the more tedious steps in the present paper. These integrals are of
the same type as the integrals in \cite{beks,egs98}, and we use  for
calculations the same methods as in those papers. First we represent
each integral as a sum of $\mu$-dependent and $\mu$-independent
integrals.  The $\mu$-independent integrals admit direct analytic
calculation. To calculate the $\mu$-dependent integrals we separate the
contributions of large and small integration momenta with the help of
an auxiliary parameter $\sigma$ such that $1\ll \sigma \ll1/\mu$. In the
region of small momenta we use the condition $\mu k\ll 1$ to simplify
the integrand, and in the region of large momenta the same goal is
achieved with the help of the condition $k\gg 1$.  Finally, for $k\sim
\sigma$ both conditions on the integration momenta hold simultaneously,
so in the sum of low-momenta and high-momenta integrals all
$\sigma$-dependent terms cancel, and we obtain a $\sigma$-independent
result for the integral (for more detailed exposition of this method
see, e.g., \cite{eksann1}). Here we skip the calculations and present
only the final result

\beq  \label{new2}
\Delta E^{II} =\biggl[- \frac{9}{8} \ln{\frac{M}{m}}
+ \frac{15}{4}\zeta{(3)} + \frac{27\pi^2}{16} +\frac{3}{2}\biggr]
\frac{\alpha(Z^2\alpha)(Z\alpha)}{\pi^3}\frac{m}{M}
~\widetilde E_F.
\eeq

\section{Total Electron Factor and the Subtracted Muon Factor}

Consider now the last contribution

\begin{equation}  \label{elfactmuonfactthree}
\Delta E^{III}=-\frac{3}{16}\frac{(Z\alpha)mM}{\pi}~\widetilde E_F
\int \frac{d^4 k}{i \pi^2k^4}
C^{(e)}_{\mu \nu}(k)C^{(\mu,b)}_{\mu \nu}(-k).
\eeq

\noindent
Due to the generalized  low energy theorem the subtracted virtual muon
Compton amplitude $C^{(\mu,b)}_{\mu \nu}(-k)$ is suppressed like
$k^2/M^2$ for momenta $k<M$ (see, e.g.,\cite{eksann2}). Hence, the
recoil correction of first order in the small mass ratio arises from
the integration region in \eq{elfactmuonfactthree} where characteristic
momenta are of order $M$. At these high integration momenta only the
leading term in the ultraviolet asymptotic expansion of the one-loop
electron factor survives in the integral.  This leading term was
calculated in \cite{eks89}, and up to the terms proportional to
$k_\mu$ and/or $k_\nu$ has the form

\beq  \label{elfactasympt}
C^{(e)}_{\mu \nu}(k)\to\frac{5\alpha}{2\pi}\frac{\gamma_{\mu}\hat {k}
\gamma_{\nu}}{k^2}.
\eeq

\noindent
Due to gauge invariance of the subtracted muon factor
$k_\mu C^{(\mu,b)}_{\mu \nu}(-k)=k_\nu C^{(\mu,b)}_{\mu
\nu}(-k)=0$, and then the terms in asymptotic expansion of the electron
factor which are linear in $k_\mu$  and/or $k_\nu$ do not give a
contribution to the energy shift in \eq{elfactmuonfactthree}. Only the
term in \eq{elfactasympt} is  relevant for the calculation of the
leading recoil correction. In addition further simplifications can be
made. The subtracted muon Compton amplitude also can be written as a
sum of terms linear in $k_\mu$ and/or $k_\nu$ and the remaining terms.
But it is easy to see that $k_\mu\gamma_{\mu}\hat
{k}\gamma_{\nu}=k^2\gamma_\mu$ has zero projection on hyperfine
splitting, and hence we can omit all terms proportional to $k_\mu$
and/or $k_\nu$ in the expression for the subtracted muon Compton
amplitude in \eq{elfactmuonfact}. Then the radiative-recoil
contribution to hyperfine splitting of order
$\alpha(Z^2\alpha)(Z\alpha)E_F$ in \eq{elfactmuonfact} reduces to

\begin{equation}
\Delta E^{III}=-\frac{3}{16}\frac{(Z\alpha)mM}{\pi}\frac{5\alpha}{2\pi}
~\widetilde E_F
\int \frac{d^4 k}{i \pi^2k^6}
\gamma_{\mu}\hat {k}\gamma_{\nu}
C^{(\mu,b)}_{\mu \nu}(-k).
\eeq

\noindent
This last integral is proportional to the integral  for
radiative-recoil corrections of order $(Z^2\alpha)(Z\alpha)(m/M)E_F$
generated by radiative insertions in the muon line \cite{eks88}.  Let
us recall that the leading term in the asymptotic expansion of the
skeleton virtual Compton amplitude is

\beq  \label{elfactasymptskel}
C^{(e,skel)}_{\mu \nu}(k)\to-2\frac{\gamma_{\mu}\hat {k}
\gamma_{\nu}}{k^2}.
\eeq

\noindent
Comparing this asymptotics with the expression in \eq{elfactasympt}
and using the result of \cite{eks88} we obtain

\beq  \label{new3}
\Delta E^{III}
=\biggl[- \frac{45}{8}\zeta{(3)}+  \frac{15 \pi^2}{4} \ln{2}
- \frac{195}{32}\biggr]\frac{\alpha (Z^2\alpha)(Z\alpha)}{\pi^3}
\frac{m}{M}~\widetilde E_F.
\eeq

\section{Summary}

The total three-loop radiative-recoil correction to hyperfine splitting
in muonium generated by the diagrams in Fig.\ \ref{radrecdiag} with
one-loop radiative photon insertions both in the electron and muon
lines is given by the sum of the contributions in \eq{new1}, \eq{new2},
and \eq{new3}

\beq \label{newrsult}
\Delta E_1 =\biggl[-\frac{15}{8}\zeta{(3)} + \frac{15\pi^2}{4}
\ln{2} + \frac{27\pi^2}{16} - \frac{147}{32} \biggr]
\frac{\alpha (Z^2\alpha)(Z\alpha)}{\pi^3}\frac{m}{M}\,
~\widetilde E_F.
\eeq

\noindent
Note that, as explained in Section \ref{logcanc}, single-logarithmic
contributions cancelled in this result. We also would like to emphasize
that unlike the case of the radiative-recoil corrections generated by
the  radiative photon insertions only in the electron or only in the
muon line, the one-loop anomalous magnetic moments of both particles
give nonvanishing contributions in \eq{newrsult}.

Some other three-loop radiative-recoil  single-logarithmic and
nonlogarithmic radiative-recoil corrections were also calculated
recently. The  corrections of order  $\alpha^2(Z\alpha)(m/M)\widetilde
E_F$ generated by the graphs with two-loop polarization insertions
(irreducible and reducible) in the two-photon exchange diagrams were
obtained in \cite{egs02}

\[
\Delta E_2= \biggl\{-\biggl[6\zeta(3)+\frac{33}{4}\biggr]\ln\frac{M}{m}
- \frac{97}{8} \zeta(3)- 16\mbox{Li}_4\biggl(\frac{1}{2}\biggr)
+\frac{2\pi^2}{3}\ln^22-\frac{2}{3} \ln^42
\]
\begin{equation}
\label{result1}
+ \frac{5\pi^4}{36}
- \frac{13\pi^2}{36}-\frac{4495}{432}\biggr\}
\frac{\alpha^2(Z\alpha)}{\pi^3}\frac{m}{M}
~\widetilde{E}_F.
\end{equation}

Single-logarithmic and nonlogarithmic corrections generated
by the diagrams with one-loop polarization insertions in the exchanged
photons and radiative photon insertions in the fermion lines were
obtained in \cite{egs03}. These are corrections of orders
$\alpha^2(Z\alpha)(m/M)\widetilde E_F$ and
$\alpha(Z^2\alpha)(Z\alpha)(m/M)\widetilde E_F$, and they have the form

\beq  \label{result2}
\Delta E_3=\biggl[\frac{22}{3}\ln{\frac{M}{m}}
+7.36110~(3)\biggr]\frac{\alpha^2
(Z\alpha)}{\pi^3}\frac{m}{M}~ \widetilde E_F
\eeq
\[
+\biggl[
\biggl(6\zeta{(3)} - 4 \pi^2 \ln{2}
+ \frac{13}{2}
\biggr)\ln{\frac{M}{m}} +22.51939(5)\biggr]
\frac{\alpha(Z^2\alpha)
(Z\alpha)}{\pi^3}\frac{m}{M}~ \widetilde E_F,
\]

Combining all three-loop single-logarithmic and nonlogarithmic
corrections to hyperfine splitting in \eq{newrsult}, \eq{result1}, and
\eq{result2} we obtain  ($Z=1$ below)

\beq
\Delta E_{tot}=\biggl[\biggl(- 4 \pi^2\ln{2}
+ \frac{67}{12}\biggr) \ln{\frac{M}{m}}
- 14\zeta(3)- 16\mbox{Li}_4\biggl(\frac{1}{2}\biggr)
+\frac{2\pi^2}{3}\ln^22
\eeq
\[
+ \frac{15\pi^2}{4} \ln{2}-\frac{2}{3} \ln^42+ \frac{5\pi^4}{36}
+ \frac{191\pi^2}{144}
-\frac{12959}{864}+29.88049~(6)\biggr]\frac{\alpha^3}{\pi^3}\frac{m}{M}
~\widetilde E_F,
\]

\noindent
or

\begin{equation}
\Delta E_{tot}=\biggl[\biggl(- 4 \pi^2\ln{2}
+ \frac{67}{12}\biggr) \ln{\frac{M}{m}}
+45.0546\biggr]
\frac{\alpha^3}{\pi^3}\frac{m}{M}
~\widetilde E_F.
\end{equation}

\noindent
Numerically this contribution to the muonium HFS is

\begin{equation}
\Delta E_{tot}=-0.019~2~\mbox{kHz}.
\end{equation}

Currently the theoretical accuracy of hyperfine splitting in muonium is
about 70 Hz. A realistic goal is to reduce this uncertainty below 10
Hz (see a more detailed discussion in \cite{egs01r,egs03}).  The new
contribution obtained in this paper, \eq{newrsult}, together with the
results of other recent research \cite{egs98,egs02,egs03,my,rh,es04}
makes achievement of this goal closer. Phenomenologically, the improved
accuracy of the theory of hyperfine splitting would  lead to
a reduction of the uncertainty of the value of the electron-muon mass
ratio derived from the experimental data \cite{lbdd} on hyperfine
splitting (see, e.g., reviews in \cite{egs01r,mt00}).

\acknowledgements

This work was supported in part by the NSF grant PHY-0138210.  The work
of V.  A. Shelyuto was also supported in part by the RFBR grants
03-02-04029 and 03-02-16843 and DFG grant GZ 436 RUS 113/769/0-1.

\end{document}